\documentclass[prl,twocolumn,superscriptaddress]{revtex4}
\usepackage{graphicx}
\usepackage{dcolumn}
\usepackage{amsmath,amsthm,amssymb}
\usepackage{subfigure}

\begin{document}

\title{Globally regular instability of $AdS_3$}

\author{Piotr Bizo\'n} \affiliation{Institute of Physics,
  Jagiellonian University, Krak\'ow, Poland}
  \affiliation{Max Planck Institute for Gravitational Physics (Albert Einstein Institute),
Golm, Germany}
\author{Joanna Ja\l mu\.zna} \affiliation{Faculty of Mathematics and Computer Science,
  Jagiellonian University, Krak\'ow, Poland}

\date{\today}
\begin{abstract}
  We consider three-dimensional AdS gravity minimally coupled to a massless scalar field and study numerically the evolution of small smooth circularly symmetric perturbations of the $AdS_3$ spacetime.   As in higher dimensions, for a large class of perturbations, we observe a turbulent cascade of energy to high frequencies which entails instability of  $AdS_3$. However, in contrast to higher dimensions, the cascade cannot be terminated by  black hole formation because small perturbations have energy below the black hole threshold. This situation appears to be challenging for the cosmic censor.  Analysing the energy spectrum of the cascade we determine the width $\rho(t)$ of the analyticity strip  of  solutions in the complex spatial plane and argue by extrapolation that $\rho(t)$ does not vanish in finite time. This provides evidence that the turbulence is too weak to produce a naked singularity and the solutions remain globally regular in time, in accordance with the cosmic censorship hypothesis.
  \begin{center}
  \emph{Dedicated to Professor Andrzej Staruszkiewicz on the occasion of the \\50th anniversary of his pioneering work on three-dimensional gravity \cite{star}.}
  \end{center}
\end{abstract}

\maketitle
\emph{Introduction.} Asymptotically AdS spacetimes  \nolinebreak have come to play a central role in theoretical
 physics, prominently
 due to the celebrated AdS/CFT correspondence which conjectures a gauge/gravity duality. By the positive energy theorem,  AdS spacetime is a ground state among asymptotically AdS spacetimes, much as Minkowski spacetime is a ground state among asymptotically flat spacetimes. However, the evolutions of small perturbations of these ground states are very different. In the case of Minkowski, small perturbations disperse to infinity and the spacetime is asymptotically stable \cite{ck}. In contrast, asymptotic stability of AdS is precluded because the conformal boundary acts like a mirror at which perturbations  propagating outwards bounce off and return to the bulk. This gives rise to complex nonlinear wave interactions, understanding of which is the key to the problem of stability of AdS spacetime.

In our recent joint work with Rostworowski \cite{br,jrb}
on spherically symmetric massless scalar field minimally coupled to AdS gravity in  dimensions $D\geq 4$ we gave numerical and perturbative evidence for the instability of $AdS_D$. More precisely, we showed that there is a large class of arbitrarily small
perturbations of $AdS_D$ that evolve into a black hole after a time of order $\mathcal{O}(\varepsilon^{-2})$, where $\varepsilon$
measures the size of a perturbation. On the basis of nonlinear perturbation analysis, we
conjectured that this instability is due to a resonant transfer
of energy from low to high frequencies, or equivalently, from coarse to fine spatial
scales, until eventually an apparent horizon forms. This mechanism is reminiscent
of the turbulent energy cascade in fluids (with black hole formation being the analogue of the viscous cut-off).

  Further studies of the same model confirmed and extended our findings \cite{bll} and provided  important new insights concerning the existence of  time-periodic solutions \cite{mr} and the transition between turbulent and non-turbulent regimes \cite{bll2}.
  The coexistence of turbulent and time-periodic solutions (geons) was also demonstrated for the vacuum Einstein equations (without any symmetry assumptions) using nonlinear perturbation analysis \cite{dhs}.

 In this paper we consider the problem of stability of $AdS_3$. The salient feature of AdS gravity in three dimensions is a  mass gap between $AdS_3$ and the lightest black hole solution, whose mass thereby provides the threshold for black hole formation (we recall that there is no such threshold in $D\geq 4$).  Since small perturbations have energy below this threshold, they cannot evolve into black holes, thus we are left with the dichotomy: naked singularity formation  or global-in-time regularity. Resolving this dichotomy numerically is challenging because in three dimensions the rate of the transfer of energy to high frequencies is much faster than in higher dimensions, which puts stringent demands on the spatial resolution and severely limits the times accessible in  simulations. Below, we will get evidence  against singularity formation  by employing the so called analyticity strip method introduced by Sulem et al. \cite{ssf}. This  method makes use of the fact that a real singularity does not come out of the blue but emerges when a complex singularity hits the real axis. Thus, tracing the motion of complex singularities and showing that they never touch the real axis provides evidence for global-in-time regularity.
\vskip 0.1cm
\emph{Model.} As in \cite{br,jrb}, we investigate the problem of stability of $AdS_3$ within
 the Einstein-scalar field  model
 \begin{equation}\label{einstein-scalar}
G_{\alpha\beta}+\Lambda g_{\alpha\beta}= \kappa T_{\alpha\beta}\,,\quad  g^{\alpha\beta} \nabla_{\alpha}
\nabla_{\beta} \phi=0\,,
   \end{equation}
 where $T_{\alpha\beta}=
 \partial_{\alpha} \phi \partial_{\beta} \phi-\frac{1}{2} g_{\alpha\beta}(\partial \phi)^2$ is the stress-energy tensor of the scalar field and $\Lambda<0$ is the cosmological constant.
Assuming the circular symmetry for the scalar field $\phi=\phi(\tau,r)$ and  the metric
\begin{equation}\label{ansatz}
  ds^2 = -\tilde A e^{-2\delta}\, d\tau^2 + {\tilde A}^{-1} dr^2 + r^2\, d\varphi^2\,,
\end{equation}
where $\tilde A$ and $\delta$ are functions of $(\tau,r)$ we get the wave equation
\begin{equation}\label{weq}
  \partial_{\tau} \left({\tilde A}^{-1} e^{\delta} \partial_{\tau} \phi\right)=\frac{1}{r}\partial_r\left(r \tilde A e^{-\delta} \partial_r \phi\right)\,,
\end{equation}
and the Einstein equations
\begin{eqnarray}
 \partial_r \delta &=& -\kappa\, r \,S\,, \label{slicing} \\
  \partial_r \tilde A  &=& -\kappa \,r \tilde A\, S - 2 \Lambda r, \label{hamilton0}\\
   \partial_{\tau} \tilde A &=& -2 \kappa\,  r \,\tilde A \,\partial_{
\tau} \phi\, \partial_r \phi\,, \label{momentum}
\end{eqnarray}
where $S={\tilde A}^2 e^{-2\delta} (\partial_t \phi)^2+(\partial_r \phi)^2$. For $\phi=0$ (vacuum), these equations have a one-parameter family of static solutions $\delta=0, \tilde A=1-M+r^2/\ell^2$, where $\ell=(-\Lambda)^{-1/2}$ is the length scale and $M\geq 0$ is the total mass. This family includes the pure AdS space for $M=0$  and BTZ black holes  for  $M\geq 1$ \cite{btz}. For $0<M<1$ the solutions have a conical singularity akin to the presence of a point mass at the origin \cite{star}.
\vskip 0.1cm
\noindent\emph{Remark.}
If $\Lambda=0$ then  it follows from Eqs.\eqref{slicing} and \eqref{hamilton0}  and the boundary conditions $\tilde A(\tau,0)=1$ and $\delta(\tau,0)=0$ that $\tilde A e^{-\delta} \equiv 1$, hence Eq.\eqref{weq} reduces to the  radial wave equation in  flat spacetime. For any finite-energy solution of this equation one can  integrate \eqref{hamilton0} to get
\begin{equation}\label{A}
  \tilde A(\tau,r)=\exp\left(-\kappa \int_0^r S(\tau,r') r' dr'\right),
\end{equation}
thus $\tilde A(\tau,r)$ is   bounded away from zero for all times. This reflects a well-known fact that in three-dimensional gravity with $\Lambda=0$ and matter satisfying the dominant energy condition there are no trapped surfaces \cite{ida}.
\vskip 0.1cm
 It is convenient to define dimensionless coordinates $(t,x)\in (-\infty,\infty)\times [0,\pi/2)$ by  $\tau=\ell t$ and $r=\ell \tan{x}$. In terms of these coordinates and  $A=(1+r^2/\ell^2)^{-1} \tilde A$, the metric \eqref{ansatz} takes the form
\begin{equation}\label{metric}
  ds^2=\frac{\ell^2}{\cos^2{\!x}} \left(-A e^{-2\delta} dt^2 + A^{-1} dx^2 + \sin^2{\!x}\, d\varphi^2\right)\,.
\end{equation}
In the following we denote derivatives with respect to $t$ and $x$ by overdots and primes, respectively, and
define auxiliary variables $\Phi=\phi'$ and $\Pi=A^{-1} e^{\delta} \dot \phi$.
We use a unit of mass such that $\kappa=1$ and a unit of length such that $\ell=1$. Then, the wave equation \eqref{weq} expressed in the first order form reads
\begin{equation}\label{weq2}
  \dot \Phi = ( A e^{-\delta} \Pi)',\quad \dot \Pi = \frac{1}{\tan{x}} (\tan{x} A e^{-\delta} \Phi)'\,.
\end{equation}
and the system (4-6) becomes
\begin{eqnarray}
 \delta' &=& -\sin{x} \cos{x} \, (\Pi^2+\Phi^2)\,, \label{delta-x}\\
  A' &=& - \sin{x} \cos{x}\, A (\Pi^2+\Phi^2) +2 \tan{x}\, (1-A), \label{hamilton}\\
   \dot A &=& -2 \sin{x} \cos{x}\,  e^{-\delta} A^2 \Pi \Phi\,. \label{momentum-x}
\end{eqnarray}
We require solutions to be smooth. This  implies that near $x=0$
 the fields behave as follows
   \begin{align}\label{bc0}
   \phi(t,x)&=f_0(t)+\mathcal{O}(x^2),\quad  \delta(t,x)=\mathcal{O}(x^2), \nonumber \\
   A(t,x) &=1+\mathcal{O}(x^2),
    \end{align}
    where we used the normalization $\delta(t,0)=0$ so that $t$ be the proper time at the center. Near spatial infinity we assume that (using $\rho=\pi/2-x$)
 \begin{align}\label{bcinf}
    \phi(t,x)&=f_{\infty}(t) \rho^2+\mathcal{O}(\rho^4), \quad \delta(t,x)=\delta_{\infty}(t)+\mathcal{O}(\rho^4),\nonumber \\ A(t,x)&=1-M \rho^2 + \mathcal{O}(\rho^4),
 \end{align}
where the power series are uniquely determined by a constant $M$ and functions $f_{\infty}(t)$ and $\delta_{\infty}(t)$. These boundary conditions ensure that the mass function defined by $m(t,x):=(1-A)/\cos^2{\!x}$ has a finite time-independent limit $M=\lim_{x\rightarrow \pi/2} m(t,x)$.
\vskip 0.1cm
\emph{Spectral properties.} It follows from  Eq.\eqref{weq2} that linear perturbations of $AdS_3$ ($\phi=0,\delta=0,A=1$) are governed by the self-adjoint operator $L=-(\tan{x})^{-1}\, \partial_x (\tan{x} \, \partial_x)$ on the Hilbert space $L^2([0,\pi/2], \tan{x}\, dx)$. The spectrum of $L$  is  $\omega_k^2=(2+2k)^2$ ($k=0,1,\dots$) and the corresponding  orthonormal eigenfunctions are given by the Jacobi polynomials
  $e_k(x)=2 \sqrt{k+1} \, \cos^2{x}\, P_k^{0,1}(\cos{2x})$.
We shall denote the $L^2$-inner product by $(f,g):=\int_0^{\pi/2} f(x) g(x) \tan{x}\, dx$.
 To analyse the spectral decomposition of solutions we define projections $\Phi_k:=(\sqrt{A}\,\Phi,e'_k)$ and $\Pi_k:=(\sqrt{A}\,\Pi,e_k)$. Then, using Eq.\eqref{hamilton} and the orthogonality relation $(e'_j,e'_k)=\omega_k^2 \delta_{jk}$ we can express the total mass as the Parseval sum
\begin{equation}\label{parseval}
    M=\frac{1}{2}\int_0^{\pi/2} \left(A\Phi^2+A\Pi^2\right) \tan{x}\,dx=\sum_{k=0}^{\infty} E_k(t)\,,
\end{equation}
where $E_k:=\Pi_k^2+\omega_k^{-2} \Phi_k^2$ can be interpreted as the energy of the $k$-th mode.
\vskip 0.1cm
\emph{Methods.}
We solve numerically the system \eqref{weq2}-\eqref{hamilton} with the boundary conditions \eqref{bc0} and \eqref{bcinf}. We use the standard method of lines with the fourth-order Runge-Kutta time integration and fourth-order spatial finite differences. The Kreiss-Oliger dissipation is added explicitely to eliminate high-frequency instabilities. The scheme is fully constrained, that is the metric functions $A$ and $\delta$ are updated at each time step by solving the slicing condition \eqref{delta-x} and the hamiltonian constraint \eqref{hamilton}. The momentum constraint \eqref{momentum-x} is only monitored to verify the accuracy of computations.
 To improve the balance between the spatial resolution and duration of simulations, we refine the entire spatial grid when a global spatial error exceeds some prescribed tolerance level. This method is computationally expensive but very stable. We usually start on a grid with $2^{12}$ points and allow for four levels of refinement.

In order to extract an information about regularity of solutions  from numerical data, we shall use the analyticity strip method \cite{ssf}. This method is based on the following idea. Consider a solution $u(t,x)$ of some nonlinear evolution equation for real-analytic initial data and let $u(t,z)$ be its analytic extension to the complex plane of the spatial variable. Typically, $u(t,z)$ will have complex singularities (coming in complex-conjugate pairs) which move in time. If a complex singularity hits the real axis, the solution becomes singular in the real world.
The imaginary part of the complex singularity $z=x+i \rho$ closest to the real axis measures the width of the analyticity strip around the real axis. Thus, monitoring the time evolution of $\rho(t)$ and checking if it vanishes (or not) in a finite time, one can predict (or exclude) the blowup. The key observation is that the value of $\rho$ is encoded in  the asymptotic behaviour of  Fourier coefficients of $u(t,x)$ which decay exponentially as $\exp(-\rho k)$ for large~$k$ (with an algebraic prefactor depending on the type of the singularity), see e.g. \cite{ckp}. Therefore, $\rho(t)$ can be obtained by fitting an exponential decay to the tail of the numerically computed Fourier spectrum.
 \vskip 0.1cm
 \emph{Results.} We solved Eqs.\eqref{weq2}-\eqref{hamilton} for a variety of small initial data. By small we mean that the total mass $M\ll 1$ (note that in 3D gravity the mass $M$ is scale invariant). For typical initial data the dynamics is turbulent. The heuristic explanation of the mechanism which triggers the turbulent behavior is the same as in higher dimensions, namely the generation of secular terms by four-wave resonant interactions \cite{br,jrb}, so we shall not elaborate on this here.  For the sake of completeness, we mention that for some solutions the mechanism of transferring energy to high frequencies is not active which is probably due to the fact that their initial data belong to stability islands around time-periodic solutions \cite{mr}.  Such non-turbulent solutions need not concern us here.

 \begin{figure} [h]
 \includegraphics[width=0.35\textwidth,angle=270]{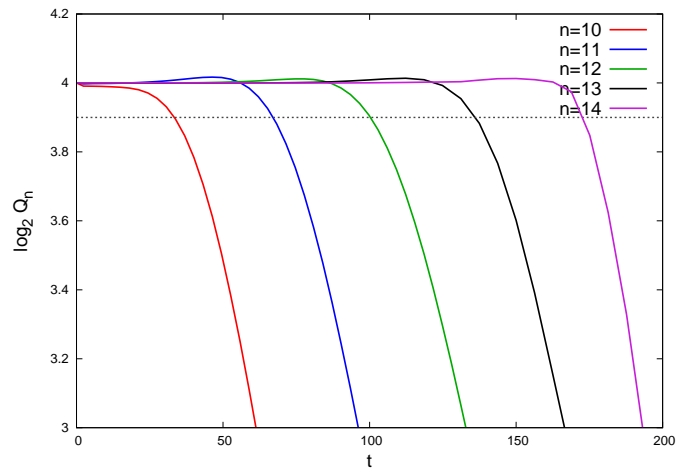}
  \caption{Results of convergence tests from runs performed on grids of size $2^n$ for $n$ from 10 to 16 (for the initial data \eqref{data} with $\varepsilon=0.3$).
  The convergence factor for the solution $\Phi_n$ computed on the $2^n$-grid is defined by
  $Q_n=\frac{||\Phi_n-\Phi_{n+1}||}{||\Phi_{n+1}-\Phi_{n+2}||}$, where $||\cdot||$ is the spatial $\ell_2$-norm.
  By convention, we define the reliability time  for the run on the $2^n$-grid as the time when $Q_n$ deviates from the expected value $2^4$ by, say, $7\%$ (depicted by the horizontal dashed line). We find empirically that the reliability time scales linearly with the product $n \,\varepsilon^{-2}$.
  }
 \label{conv}
\end{figure}

  The results presented below were generated from the time-symmetric Gaussian initial data of the form
  \begin{equation}\label{data}
  \phi(0,x)=\varepsilon \exp(-\tan^2{\!x}/\sigma^2)\,,\quad \dot\phi(0,x)=0
  \end{equation}
   with width $\sigma=1/32$ and varying small amplitudes $\varepsilon$.
 For these data, the evidence for the expected fourth-order convergence is shown in Fig.~1. The loss of convergence (which is inevitable in numerical simulations of turbulent phenomena) is clearly apparent beyond some 'reliability time'  signalling that the small scales become unresolved. We estimate that the reliability time corresponding to the smallest amplitude $\varepsilon=0.3$ and the highest resolution $2^{16}$ used in our simulations is about $230$ and, consequently, we did not use any longer-time data in our analysis.

A quantity of fundamental interest for the understanding of turbulent dynamics is that of
  the energy spectrum, that is the distribution of the total energy over the modes of
the linearized problem, as given in \eqref{parseval}. Fig.~2 shows how the energy spectrum develops in time. The range of frequencies participating in the evolution is seen to increase very rapidly (cf. \cite{prague} where an analogous plot of the energy spectrum in 4D is shown).

\begin{figure}[h]
 \includegraphics[width=0.35\textwidth,angle=270]{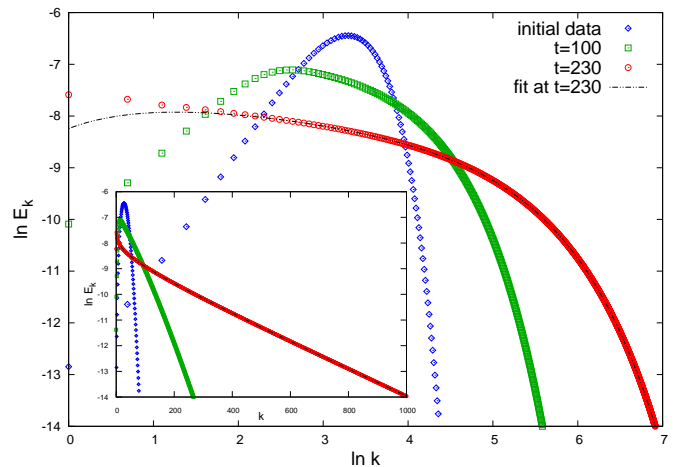}
  \caption{Energy spectra at three instants of time for the initial data \eqref{data} with $\varepsilon=0.3$ (which gives $M=0.044$). The fit of the formula (\ref{fit}) in the interval $10<k<1000$ to the data at $t=230$  is depicted by the black dotted line. The inset displays the same plot in the linear-log scale to better see the exponential decay of the tails.}
 \label{spectrum}
\end{figure}

In accord with the analyticity strip method we assume that for large wave numbers the energy spectrum is described by the formula
 \begin{equation}\label{fit}
   E_k(t)=C(t)\,k^{-\beta(t)} e^{-2 \rho(t) k}\,.
 \end{equation}
 Fitting this formula to the numerical data  at a sequence of times  we obtained the time dependence of the parameters $C(t), \beta(t)$, and $\rho(t)$. We found that the width of analyticity $\rho(t)$ stays bounded away from zero and after some transient period is well approximated by the exponential decay
 \begin{equation}\label{delta}
 \rho(t)=\rho_0 \, e^{-t/T}\,,
  \end{equation}
  where $T$ is a characteristic decay time. The evidence for \eqref{delta} is shown in Fig.~3.  We find that $\rho_0$ is approximately independent of $\varepsilon$, while $T \propto \varepsilon^{-2}$. At the reliability time $\rho(t)$ is of the order of hundred mesh sizes, reassuring us that the fitting procedure is self-consistent and credible. Despite the good quality of the fit, it would probably be premature to  extrapolate  the exact behaviour \eqref{delta} indefinitely; actually, we suspect that crossovers to faster (but still exponential) decay may occur at later times. Anyway, the results suggest that $\rho(t)$ does not vanish in a finite time
     and consequently the solutions remain smooth forever. Of course,
    higher resolution simulations would be helpful in validating or refuting this conjecture.

   \begin{figure}[h]
 \includegraphics[width=0.35\textwidth,angle=270]{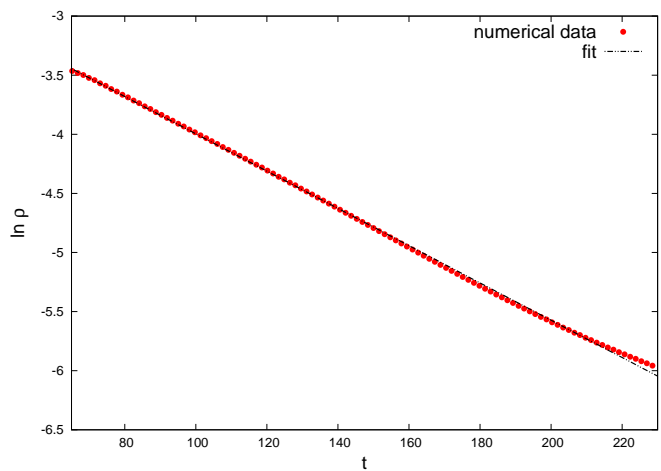}
  \caption{Time evolution of the width of analyticity $\rho$, obtained by fitting the formula (\ref{fit}) to the energy spectra for the same initial data as in Fig.~2. The result is fairly insensitive to the choice of the fitting interval. The fit of the formula \eqref{delta}  for $70<t<230$  gives $\rho_0=0.09$ and $T=63.4$ (dashed line).}
 \label{log_dec}
\end{figure}

The exponentially fast shrinking of the width of analyticity is reflected in the exponentially fast growth of higher Sobolev norms $H_s$ with $s>1$, implying the instability of $AdS_3$. This is illustrated  in Fig.~4 which depicts the time evolution of the second homogenous Sobolev norm $\dot H_2$. After an initial quiescent period, whose duration scales as $\varepsilon^{-2}$, the maxima of $\dot H_2(t)$ begin to grow exponentially  approximately as  $\exp(t/T)$ (as could be guessed from \eqref{delta} by dimensional analysis).

\begin{figure} [h]
 \includegraphics[width=0.35\textwidth,angle=270]{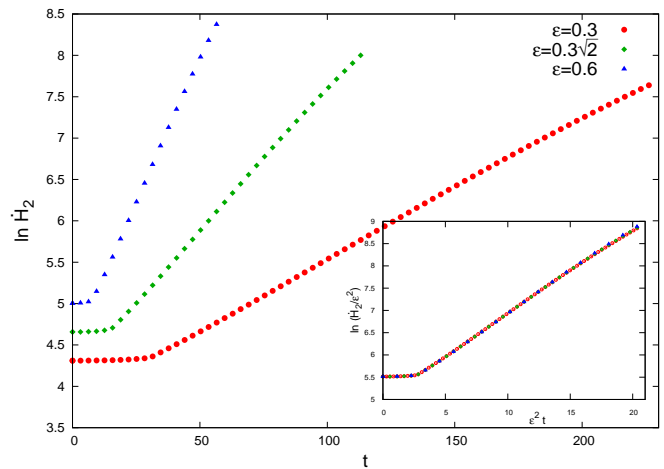}
  \caption{Time evolution of the $L^2$-norm of the second spatial derivative $\dot H_2=||\phi''(t,x)||_2 $. This quantity rapidly oscillates in time so for clarity only the upper envelope of oscillations is plotted. In the inset,  the curves corresponding to three different amplitudes $\varepsilon$ are shown to coincide after rescaling.}
 \label{norm}
\end{figure}
\vskip 0.1cm
\emph{Conclusions.}
The numerical results presented above indicate that initially smooth small perturbations of $AdS_3$ remain smooth forever, however they
do not remain small in any reasonable norm that captures the turbulent behavior, hence $AdS_3$ is unstable.
This kind of gradual loss of regularity, where  solutions develop progressively finer spatial scales as $t\rightarrow \infty$ without ever losing smoothness (sometimes referred to as weak turbulence), has been  well known in fluid dynamics, for example it has been proved for the incompressible Euler equation in two spatial dimensions \cite{yu, bbz}. More recent studies show  that weak turbulence is common for nonlinear wave equations in bounded domains, see e.g. \cite{ckstt, gg,cf,gt}. We point out that in the case of Einstein's equations,  the weakly turbulent dynamics can proceed forever only in 3D, whereas in higher dimensions it is unavoidably cut off in finite time by the black hole formation.

 Above we discussed only small mass solutions but we observed a very similar behavior in the whole range of masses
 $0<M<1$ so it is tempting to conjecture  that all solutions with $M<1$ are globally regular in time.
 Such a finite energy threshold for blowup is typical for nonlinear wave equations in critical dimensions where a finite amount of energy must concentrate to produce a singularity, a notable example being $3D$ wave maps  \cite{st}. Finally, let us mention that the threshold at $M=1$ was investigated numerically in \cite{pc} leading to important insights about the near-critical dynamics, however the critical solution itself remains not understood \cite{carsten}.
\vskip 0.1cm
 \noindent \emph{Acknowledgments:}
We thank Andrzej Rostworowski for helpful remarks. PB gratefully acknowledges enlightening conversations with  Vladim\'ir \v{S}ver\'ak and Edriss Titi on turbulence. This work was initiated when PB visited the Banff International Research Station (BIRS), Canada, and completed when JJ visited the Albert Einstein Institute (AEI) in Golm. We thank BIRS and AEI for their warm hospitality. This work was supported by the NCN grant DEC-2012/06/A/ST2/00397 and  Foundation for Polish Science under the MPD
Programme 'Geometry and Topology in Physical Models'. The computations were performed at the Academic Computer Centre Cyfronet AGH using the PL-Grid infrastructure.

\end{document}